\documentclass[conference]{IEEEtran}

\IEEEoverridecommandlockouts

\usepackage{amsmath}   
\usepackage{amssymb}
\usepackage{mathrsfs}

\usepackage{balance}   
\usepackage{flushend}

\usepackage{cite}      

\usepackage{ifpdf}     

\usepackage{multicol}
\usepackage{makeidx}
\usepackage{color}
\usepackage{pseudocode}
\usepackage{epsf}
\usepackage{subfigure}

\usepackage{comment}

\ifCLASSINFOpdf
   \usepackage[pdftex]{graphicx}
   \graphicspath{{../fig/pdf/}}
   \DeclareGraphicsExtensions{{.pdf}}
\else
   \usepackage[dvips]{graphicx}
   \graphicspath{{../fig/eps/}}
   \DeclareGraphicsExtensions{{.eps}}
\fi

\includecomment{nonDraftFigure} 
\excludecomment{draftFigure}

\makeatletter
\def\normalsize{\@setfontsize{\normalsize}{10bp}{10.00pt}}
\normalsize
\makeatother

\DeclareMathAlphabet{\mathpzc}{OT1}{pzc}{m}{it}

\newcommand{\be}{\begin{eqnarray}}
\newcommand{\ee}{\end{eqnarray}}

\begin{document}

\title{LDPC Code Design for \\ Noncoherent Physical Layer Network Coding \vspace{-0.15cm} }

\author{
Terry Ferrett and
Matthew C. Valenti,\\
West Virginia University, Morgantown, WV, USA.
\vspace{0cm}
}

\maketitle

\begin{abstract}

This work considers optimizing LDPC codes
in the physical-layer network coded two-way relay channel
using noncoherent FSK modulation.
The error-rate performance of channel decoding at the relay node 
during the multiple-access phase was improved through EXIT-based 
optimization of Tanner graph variable node degree distributions.
Codes drawn from the DVB-S2 and WiMAX standards were
used as a basis for design and performance comparison.
The computational complexity characteristics of the standard
codes were preserved in the optimized codes by maintaining
the extended irregular repeat-accumulate (eIRA).
The relay receiver performance was optimized considering two modulation orders
 $M=\{4, 8\}$ using iterative decoding in which the decoder and demodulator
refine channel estimates by exchanging information.
The code optimization procedure yielded unique optimized
codes for each case of modulation order and available channel
state information.
Performance of the standard and optimized codes were measured
using Monte Carlo simulation in the flat Rayleigh
fading channel, and error rate improvements
up to $1.2$ dB are demonstrated depending on system
parameters.

\end{abstract}

\IEEEpeerreviewmaketitle


\section{Introduction}
The \emph{two-way relay channel} (TWRC) models two \emph{source nodes} outside 
radio range exchanging information through a single \emph{relay node} in range of both.
\emph{Physical-layer network coding} (PNC) \cite{zhang2:2006} may applied in the TWRC
to increase throughput by reducing the number of time slots
required for exchange.
The exchange between the sources is broken into
two phases: the \emph{multiple-access} (MA) phase and \emph{broadcast} phase.
During MA, the sources transmit at the same time to the relay,
and the signal received at the relay is the electromagnetic
sum of signals transmitted by the sources.
During the broadcast phase, the relay transmits the network-coded
sum of signals to the tranceivers, and each performs network
decoding to recover the information transmitted by the opposite sources.

Previous work developed a demodulation and decoding
scheme for the relay in the PNC-coded TWRC using 
\emph{M-ary FSK modulation} \cite{vtf:2011} \cite{ferrett:2013}.
This scheme supports iterative decoding between decoder
and demodulator to improve error-rate performance of the received
network-coded information bits.
The receiver supports power-of-two modulation orders and improved decoding performance 
based on available channel state information (CSI).
A simulation regime was performed to demonstrate the performance
of the scheme using Turbo decoding at the relay.

The error-rate performance of LDPC codes depends on the
structure of the parity check matrix and thus the properties of the Tanner Graph
and degree distribution.
In this work we apply an optimization technique which identifies
\emph{variable node degree distributions} likely to yield good codes
using extrinsic information tranfer characteristic (EXIT) curve fits \cite{brink:2004}. 
LDPC codes taken from the \emph{DVB-S2} and \emph{WiMAX} standards are used as a basis
for design and performance comparison, and their variable node degree distributions
are optimized to yield codes exhibiting error-rate performance
superior to the standard codes.
The complexity characteristics of the standard codes are maintained
in the optimized codes by preserving the \emph{extended irregular repeat-accumulate}
constraint.

A variety of optimization techniques exist for improving
LDPC code performance.  
Random permutation matrices, large girth optimization,
and column weight optimization have been applied
to WiMax standardized codes yielding  
gains up to $0.4$ dB \cite{peng:2009}.
Joint design of and LDPC codes under carrier frequency offsets demonstrate the relationship
between system properties and code performance \cite{wenwen:2012}.

A PNC scheme compensating for
correlation among transmitted symbols introduced
by channel coding to improve decoder performance at the relay
has been considered \cite{zhang:2011}.
In this work it is demonstrated that LDPC codes
having different degree distributions are preferred in different
SNR regimes. Roughly, smaller degrees are preferable at low SNR
and higher degrees at high SNR.
A PNC system used as a basis for optimizing LDPC codes via
EXIT analysis demonstrated performance $1.5$ dB from capacity \cite{li:2013}.


The rest of this work is organized as follows.
Section II develops the system model applied for code optimization
and error-rate simulation.
Section III describes the variable node degree distribution optimization technique in information-theoretic terms,
while Section IV describes the computational procedure followed to generate the optimized codes
and demonstrates the performance of the codes versus standard.
Section V provides concluding remarks.

\section{System Model}\label{sec:SystemModel}

\begin{figure}[t]
  \centering
\vspace{-5mm}
  \includegraphics[width=9cm]{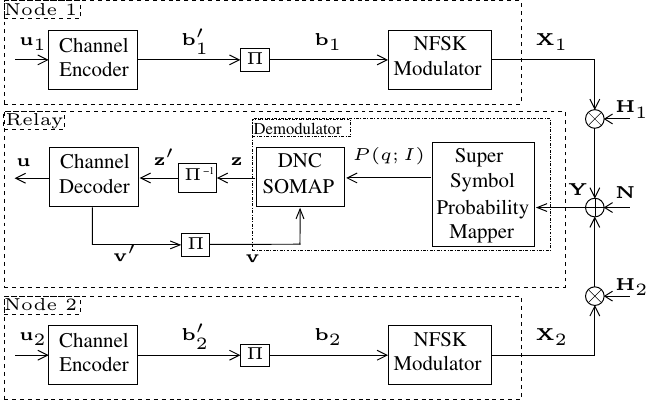}
  \caption{System Model - TWRC DNC MAC Phase}
  \label{fig:sysm}
\end{figure}

This section describes the model applied for LDPC code optimization and error-rate simulation.
The model is depicted in Figure \ref{fig:sysm}.

\subsection{Transmission by Source Nodes}

The \emph{source} nodes $\mathcal{N}_i, i \in \{1, 2\}$ generate binary information sequences $\mathbf{u}_i = [ u_{1,i}, ..., u_{K,i}]$ having length $K$.
A rate-$R$ LDPC code is applied to each $\mathbf{u}_i$, generating a length $N = K/R$ channel codeword, denoted by $\mathbf{b}'_i = [b_{1,i} ..., b_{L,i}]$.
The codeword is passed through an interleaver, modeled as a permutation matrix $\mathbf{\Pi}$ having dimensionality $N \times N: \mathbf{b}_i = \mathbf{b}'_i \mathbf{\Pi}$.
Let $\mathcal{D} = \{ 0, ..., M-1 \}$ denote the set of integer indices corresponding to each FSK tone, where $M$ is the modulation order.
The number of bits per symbol is $\mu = \log_2 M$.
The codeword $\mathbf{b}_i$ at each node is divided into $N_q = N/\mu$ sets of bits, each of which is mapped to an $M$-ary symbol $q_{k,i} \in \mathcal{D}$, where $k$ denotes the symbol number, and $i$ denotes the source.

The modulated signal transmitted by source $\mathcal{N}_i$ during signaling interval $k T_s \leq t < (k+1) T_s $ is

\vspace{-4mm}
\begin{align}\label{eqn:td_sym}
  s_{k,i}(t) =  \sqrt{ \frac{2}{T_s} }
  \cos
  \left[
    2 \pi
    \left(
      f_c
      + \frac{q_{k,i}}{T_s}
    \right )
    (t - kT_s)
  \right]
\end{align}
\vspace{-3mm}

\noindent where $s_{k,i}(t)$ is the transmitted signal, $f_c$ is the carrier frequency, and $T_s$ is the symbol period.

The continuous-time signals $s_{k,i}(t)$ are represented in discrete-time by the set of column vectors $\{ \mathbf{e}_{q_k,i}: q_{k,i} \ \in \ \mathcal{D}\}$.
The column vector $\mathbf{e}_{q_{k,i}}$ is length $M$, contains a $1$ at vector position $q_{k,i}$, and $0$ elsewhere.
The modulated codeword from source $\mathcal{N}_i$ is represented by the matrix of symbols $\mathbf{X}_i = [\mathbf{x}_{1,i},...,\mathbf{x}_{N_q,i}]$, having dimensionality $M \times N_q$, where $\mathbf{x}_{k,i} = \mathbf{e}_{q_{k,i}}$.

\subsection{Channel Model}

All channels are modeled as flat-fading channels having independent gains for every symbol period.
The gain from node $\mathcal{N}_i$ to the relay during a particular signaling interval $k$ is denoted by $h_{k,i}$.
The gain is represented as $h_{k,i} = \alpha_{k,i} e^{j \theta_{k,i}}$, where $\alpha_{k,i}$ is the received amplitude and $\theta_{k,i}$ is the phase.  The received amplitude has a Rayleigh distribution with parameter $\sigma = \sqrt{\frac{1}{2}}$, and $\theta_{k,i}$ is the phase having a Uniform distribution over interval $\left[0, 2\pi \right)$.



Consider transmission of a single frame of $N_q$ symbols to the relay.
The received frame is

\vspace{-3mm}
\begin{align} \label{eqn:rec_sym}
  \mathbf{Y} = \mathbf{X}_1 \mathbf{H}_1 + \mathbf{X}_2 \mathbf{H}_2 + \mathbf{N}
\end{align}
\vspace{-3mm}

\noindent where $\mathbf{H}_i$ is an $N_q \times N_q$ diagonal matrix of channel coefficients having value $h_{k,i}$ at matrix entry $(n,n), n \in \{1,2,...,N_q\}$ and $0$ elsewhere, and $\mathbf{N}$ is an $M \times N_q$ noise matrix.
A single column of $\mathbf{Y}$ represents a single signaling interval, is denoted by $\mathbf{y}$, and referred to as a \emph{channel observation}. In terms of this definition, $\mathbf{Y} =  [\mathbf{y}_{1}, ..., \mathbf{y}_{N_q}]$, where $\mathbf{y}_k$ denotes the $k$-th channel observation. 
Denote the $k$-th column of $\mathbf{N}$ by $\mathbf{n}_k$.
Each column is composed of zero-mean circularly symmetric complex Gaussian random variables having covariance matrix $N_0 \mathbf{I}_M$; i.e., $\mathbf{n}_k \sim \mathcal{N}_c(\mathbf{0}, N_0 \mathbf{I}_M)$.
$N_0$ is the one-sided noise spectral density, and $\mathbf{I}_M$ is the $M$-by-$M$ identity matrix.
The noise spectral density is defined as $N_0 = \frac{1}{10^{X/10} R M}$, where $X$ is the bit-energy to noise ratio in decibels $\frac{E_b}{N_0}$.
Increasing modulation order decreases the noise spectral density to reflect an increase in energy per-symbol when utilizing constant energy per-bit.

\subsection{Relay Reception}

The goal of the digital-network network coding (DNC) relay receiver is to detect the network-coded combination of information bits transmitted by the end nodes, $\mathbf{u} = \mathbf{u}_1 \oplus \mathbf{u}_2$.
The relay receiver takes as input the frame of channel observations $\mathbf{Y}$.
The symbols transmitted by the sources are assumed to be received in perfect synchronization at the receiver.
Demodulation and iterative channel-decoding are applied to detect $\mathbf{u}$.
Define the network codeword as

\begin{align}
  \mathbf{b} &= \mathbf{b}_1 \oplus \mathbf{b}_2 \nonumber \\
  &=[\ b_0(\mathbf{x}_{k,1}) \oplus b_0(\mathbf{x}_{k,2}) \ ... \ b_{\mu- 1}(\mathbf{x}_{k,1}) \oplus b_{\mu-1}(\mathbf{x}_{k,2}) \  ] \nonumber \\
  & \forall \ k \in \{ 1, ..., N_q\}
\end{align}

\noindent where $b_m(\mathbf{x}_{k,i})$ denotes the $m$-th bit mapped to the $k$-th symbol transmitted by source $\mathcal{N}_i$.
Since the channel code considered is a systematic linear code, $\mathbf{b}$ forms a code from the codebooks used by the source nodes, thus, the channel decoding operation yields a hard decision on the network coded message bits $\mathbf{u}$.

The demodulator is implemented using \emph{super-symbol probability mapping} and \emph{DNC soft bit-mapping} (SOMAP),
having formulation described in \cite{ferrett:2013}.

The super-symbol probability mapper takes as input the matrix of received symbols $\mathbf{Y}$ and produces estimates
of the probability of receiving each super symbol $q_k$, defined as the tuple

\vspace{-3mm}
\begin{align}                                                                                                                       
  q_k = (q_{k,1},q_{k,2}) \ \ \ q_{k,1},q_{k,2} \in \mathcal{D} \ \ \ q_k \in \mathcal{D} \times \mathcal{D}
\end{align}
\vspace{-3mm}

\noindent to produce estimates $P(q_k)$, $ 0 < k < N_q$.
The cardinality of $\mathcal{D} \times \mathcal{D}$ is $M^2$, thus, the number of probability estimates computed 
at the relay during each symbol period grows exponentially with modulation order.
This computational complexity is a practical consideration for implementing the relay receiver.

The super-symbol probability estimates are passed to the DNC SOMAP along with with extrinsic decoder feedback $\mathbf{v}$
to produce extrinsic information $\mathbf{z}$, likelihood ratios of the network-coded bits $\mathbf{b}$ communicated by the frame.

The extrinsic information is deinterleaved to produce $\mathbf{z}' = \mathbf{z} \mathbf{\Pi}^{-1}$ and passed to the decoder.
The decoder refines the estimate of $\mathbf{z}'$, producing new extrinsic information $\mathbf{v}'$ which is interleaved to produce $\mathbf{v} = \mathbf{v}' \mathbf{\Pi}$ and returned to the SOMAP.
After a specified number of iterations, a hard decision is made on the network coded bits $\mathbf{u}$.

The transmitted symbol energies and noise spectral density $N_0$ are known at the demodulator. 
The demodulator may operate under several cases of channel state information (CSI) \cite{ferrett:2013}: the case in which the gain are completely known (full CSI), the case in which only the fading amplitudes $\alpha_{k,i}$ are known (partial CSI), and the case in which no information about the gains is known (no CSI).

Since the focus of this work is on the relay reception phase, details of the relay-to-source broadcast phase are omitted.
Interested readers are referred to \cite{ferrett:2013}.

\section{LDPC Code Optimization}\label{sec:ldpc_opt}

This section describes the application of a procedure for optimizing LDPC code error-rate performance based on \cite{brink:2004} under a range of receiver configurations to the DNC relay reception phase by varying the LDPC code degree distribution.
Relevant LDPC concepts are introduced, followed by a description of metrics
relevant to optimization.
A theoretical description of the optimization technique is provided followed by the procedure taken.

The goal is to produce LDPC code designs
having the same rate and complexity
as off-the-shelf codes from the DVB-S2 and WiMAX standards, while improving
error rate performance through optimization of the parity check matrix. 
The codes are constrained to the class of \emph{extended
  irregular repeat accumulate} (eIRA) codes.
This constraint ensures computational efficiency by
guaranteeing systematic encoding \cite{yang:2004}.
The codes considered are \emph{check-regular}, 
meaning that every check node has the same degree, $d_c$.

An LDPC code is a \emph{linear block code} having a sparse parity check matrix $\mathbf{H}$, which is an $N-K \times N$ binary matrix.
An alternative representation for the parity check matrix is the \emph{Tanner Graph}, a bipartite graph in which
one of the sets contains \emph{variable nodes}, and the other contains \emph{check nodes} \cite{proakis:2008}.
Each column of $\mathbf{H}$ corresponds to a \emph{variable node}, and each row a \emph{check node} \cite{proakis:2008}. 
A $1$ entry in $\mathbf{H}$ at row and column $(k,n)$ corresponds to an edge between the $k$-th check node and $n$-th variable node.
The \emph{degree} of a node is its number of incident edges, thus, the number of $1$'s in each row and column for each check and variable node, respectively.

The variable and check nodes may be modeled as soft-input, soft-output
decoders denoted as VND and CND, respectively \cite{brink:2004}.
The VND decoder takes a channel LLR $L_{c,VND}$ and a-priori information $L_{a,VND}$ as input
and produces a-posteriori information $L_{o,VND}$ as output.
The extrinsic information is defined as $L_{e,VND} = L_{a,VND} - L_{o,VND}$.
The mutual information between the $L_{e,VND}$ and the corresponding network-coded bit is
$I_{E,VND}$, while the mutual information between $L_{a,VND}$ and the network-coded bit is
$I_{A,VND}$.

The CND decoder takes a-priori information $L_{a,CND}$ as input
and produces a-posteriori information $L_{o,CND}$ as output.
The extrinsic information is $L_{e,CND} = L_{a,CND} - L_{o,CND}$.
The mutual information between the $L_{e,CND}$ and the corresponding network-coded bit is
$I_{E,CND}$, while the mutual information between $L_{a,CND}$ and the network-coded bit is
$I_{A,CND}$.

An \emph{EXIT chart} visualizes the transfer characteristic of a decoder by plotting
the a-priori mutual information against the extrinsic.
It has been shown that matching the VND and CND transfer characteristics
of a particular LDPC code through selection of variable node degrees
improves error rate performance \cite{ashikhmin:2004}.

\subsection{Optimization through Selection of Variable Node Degree}\label{subsec:opt_vnd}

This section describes the analytic formulation for degree optimization.
The degree distribution is defined as the set containing the degrees assigned to the variable nodes,
and the number of variable nodes taking each particular degree.
Denote a degree distribution as $V = \{\ d_{v,1}:o_1, ...,  d_{v,D}: o_D \ \}$,
where $o_i$ is the number of variable nodes having degree $d_{v,i}$.
For a particular code, all check node degrees are the same, thus,
the code is \emph{check-regular}. 
Denote the check node degree for a particular code as $d_c$.
Define $a_i$ as the fraction of variable nodes having degree $d_{v,i}$

\vspace{-3mm}
\begin{align}\label{eqn:ai}
  a_i = \frac{o_i}{N}, \ i=1,...,D
\end{align}
\vspace{-2mm}

\noindent each taking a value $0 < a_i < 1$, and $\sum_{i=1}^D{a_i} = 1$.

A realizable degree distribution satisfies the constraints imposed by the LDPC code parameters.
The sum total of edges incident on the variable and check nodes must be equal.
The number of edges incident on variable nodes having degree $d_{v,i}$ is $e_{v,i} = a_i d_{v,i} N$,
thus, the total number of edges incident on all variable nodes is

\vspace{-5mm}
\begin{align}\label{eqn:vnd_edges}
  e_{v} = \sum_{i=1}^{D} e_{v,i} = \sum_{i=1}^{D} a_i d_{v,i} N
\end{align}
\vspace{-4mm}

\noindent 
and the total number of edges incident on the check nodes is

\vspace{-5mm}
\begin{align}\label{eqn:cnd_edges}
  e_{c} = d_{c} (N - K).
\end{align}
\vspace{-4mm}

\noindent Equating $e_v$ and $e_c$ and rearranging,

\vspace{-4mm}
\begin{align}\label{eqn:edge_const}
  \sum_{i=1}^{D} \frac{a_i d_{v,i} N}{d_c (N-K)} = 1.
\end{align}
\vspace{-4mm}

\noindent A particular degree distribution $V$ is selected by choosing parameters $V$ and $d_c$ which
satisfy (\ref{eqn:edge_const}) and the sum constraint on $a_i$.
The challenge is to select a $V$ which optimizes
error rate performance for particular channel conditions and relay receiver configurations.

The optimization procedure requires generating EXIT curves corresponding
to the variable-node decoder (VND) and check-node decoder (CND).
The VND curve is described by the variable node degree distribution $V$ and the mutual
information computed between the DNC-SOMAP output and the network coded bits.
The CND curve is completely described by the degree of the check nodes $d_c$,
and is fixed for all variable node degree distributions.

Generation of the VND curve begins by computing the mutual information
between the soft-values at the output of the DNC-SOMAP and the network-coded
bits, denoted as the \emph{detector characteristic} $I_{E,DET}(I_{A,DET}, \frac{E_b}{N_0})$.
$I_{A,DET}$ is the mutual information between the DNC-SOMAP \emph{a-priori} input
and the network-coded bits, and $\frac{E_b}{N_0}$ is the signal-to-noise ratio of the super-symbol.
The detector characteristic $I_{E,DET}$ is generated by Monte Carlo simulation
under the assumption that the \emph{a-priori} input $I_{A,DET}$ is conditionally Gaussian.
The Gaussian assumption can be verified empirically using histograms of several decoding runs \cite{valenti:2005}.

The VND curve is found by computing a separate curve for each variable node degree and
combining each to form the final curve describing the entire variable node detector.
The curve corresponding to the $i$-$th$ degree is given by

\vspace{-4mm}
\begin{align}\label{eqn:vnd_curve_one_deg}
  &I_{E,VND} ( I_{A,VND}, d_{v,i}, I_{E,DET} ) = ...\nonumber \\
  &J \left( \sqrt{ (d_{v,i} - 1) J^{-1}(I_{A,VND})^2  + [ J^{-1} (I_{E,DET}) ]^2  }\right)
\end{align}
\vspace{-3mm}

\noindent where the $J$-function is defined in \cite{brink:2001} and is computed
using the truncated-series representation presented in \cite{torrieri:2011},
and $I_{A,VND} = I_{A,DET}$.
The VND curve describing the entire detector is given by

\vspace{-4mm}
\begin{align}\label{eqn:vnd_curve_overall}
  &I_{E,VND} ( I_{A,VND}, I_{E,DET} ) = ... \nonumber \\
  &\sum_{i=1}^{D} b_i \cdot I_{E,VND} ( I_{A,VND}, d_{v,i}, I_{E,DET} )
\end{align}
\vspace{-3mm}

\noindent where $b_i = e_{v,i}/\sum_{i=1}^D e_{v,i}$ is the fraction of edges incident on variable nodes of degree
$d_{v,i}$.

The CND curve is computed according to \cite{brink:2004}

\vspace{-4mm}
\begin{align}\label{eqn:ie_cnd}
  I_{E,CND} (I_{A,CND}, d_c) = 1 - J \left( \sqrt{ d_c - 1 } \cdot J^{-1} (1 - I_{A,CND}) \right)
\end{align}
\vspace{-3mm}

\noindent where $I_{A,CND}$ is the mutual information between the check node inputs and the network-coded bits.
Note that the CND curve is independent of the particular variable node degree distribution
$V$.

The curve fit for the VND and CND mutual information metrics corresponding to a particular degree distribution is performed
by plotting the curves for a range of values of $\frac{E_b}{N_0}$ and noting the
value in which the curves barely touch \cite{brink:2004}.
Specifically, the VND curve is plotted with $I_{A,VND}$ on the horizontal axis
and $I_{E,VND}$ along the vertical.
The CND curve is plotted with $I_{E,CND}$ along the horizontal
and $I_{A,CND}$ along the vertical.
The value of $\frac{E_b}{N_0}$ is varied from highest specified point to lowest,
and the highest point for which the curves touch is defined as the \emph{EXIT threshold}.
Considering a particular relay receiver and channel configuration, the optimal variable
node degree distribution is considered as the distribution having the lowest EXIT threshold.

\section{EXIT-Optimized LDPC Code Performance}\label{sec:ldpcperf}

This section presents the error-rate performance of EXIT-optimized LDPC codes and the simulation
optimization procedure used to generate the codes.
Several cases of modulation order and relay receiver CSI are considered.
LDPC codes defined by the DVB-S2 and WiMax standards are used as reference codes.

\subsection{Optimization Procedure}

The purpose of this subsection is to describe the procedure for generating EXIT curves describing the performance of selected degree distributions.

The procedure involves determining the detector EXIT characteristic through simulation for a chosen receiver configuration and channel model, and using this characteristic to generate the combined variable node/detector and check node characteristics over a range
of SNR \cite{brink:2004}.

The notation $[\cdot]$ following a bold variable denotes a vector element, for instance, $\mathbf{Z}[k]$ denotes the $k$-th element of one-dimensional vector $\mathbf{Z}$ and $\mathbf{W}[i,k]$ denotes the $(i,k)$-th element of two-dimensional vector $\mathbf{W}$.

\subsubsection{Detector Characteristic}\label{ss:detchr}

This section describes the computation of the detector characteristic for the DNC relay receiver.
These values are computed via Monte Carlo simulation as follows

\begin{enumerate}
  
\item Select source and relay frame size $L$, modulation order $M$, received channel state information (CSI or no CSI) and SNR value.

\item Select a discrete range of a-priori demodulator mutual information values, represented as the vector $\mathbf{I}_{A,DET}[k] = k/B, 0 \le k < B$.  In all of our experiments, the value $B = 100$ provided visibly smooth detector transfer characteristics.
  
\item Simulate a frame of channel-corrupted relay-received symbols $\mathbf{Y}$ according to the model given
  in Section \ref{sec:SystemModel}, having length $N_q = L/\mu$ and modulation order $M$.  
  
\item For every value of $\mathbf{I}_{A,DET}[k]$,

  \begin{itemize}
  
  \item  Generate a vector of length $L$ a-priori LLRs $\mathbf{v}$ representing LDPC decoder feedback under the assumption that the feedback is conditionally Gaussian 
    
    \vspace{-5mm}
    \begin{align}
      \mathbf{v} = (\mathbf{b} - 1/2) \sigma_k^2 + \mathbf{x} \sigma_k
    \end{align}
    \vspace{-4mm}

    where $\sigma_k = J^{-1}(\mathbf{I}_{A,DET}[k])$ \cite{brink:2004} and $\mathbf{x}$ is a length-$L$ vector of i.i.d. samples of a standard normal distribution.
    
  \item Compute the decoder extrinsic information $\mathbf{z}$ according to the receiver formulation provided in \cite{ferrett:2013}.
    
  \item Compute the simulated value of $\mathbf{I}_{E,DET}[k]$ as \cite{valenti:2005}
    
    \vspace{-4mm}
    \begin{align}
      \mathbf{I}_{E,DET}[k] = 1 - \frac{ 1 }{ L \log{2} } \sum_{\ell=0}^{L}{ \max*(0, \mathbf{z}[\ell] (-1)^{\mathbf{b}[\ell]} ) }
     \end{align}
    \vspace{-3mm}

  \item Numerically fit a third-order polynomial to the sampled detector characteristic using $\mathbf{I}_{A,DET}$ as the abscissa and $\mathbf{I}_{E,DET}$ as the ordinate, yielding coefficients $f_0, ..., f_3$

    \vspace{-4mm}
    \begin{align}
      \mathbf{I}_{E,DET}[k] =& f_3 {\mathbf{I}_{A,DET}[k]}^{3} + f_2 {\mathbf{I}_{A,DET}[k]}^{2} + \nonumber \\
      &f_1 \mathbf{I}_{A,DET}[k] + f_0.
    \end{align}
    \vspace{-3mm}    

    Define function $f_{DET}(x) = f_3x^3 + f_2x^2 + f_1x + f_0$.
    This function approximates the detector characteristic and is used in the computation
    of the combined detector/decoder characteristic.

  \end{itemize}

\end{enumerate}

\subsubsection{Combined Detector/Decoder Characteristic}\label{ss:decchr}

This subsection describes the simulation procedure for generating the combined detector/decoder characteristics.

\begin{enumerate}

\item Select a discrete range of a-priori demodulator mutual information values, represented as the vector $\mathbf{I}_{A,VND}[k] = k/B, 0 \le k < B$.  In all of our experiments, the value $B = 100$ provided visibly smooth detector transfer characteristics.

\item For every value of detector a-priori mutual information $\mathbf{I}_{A,VND}$,

  \begin{itemize}

  \item For every degree $d_{v,i}$ compute the a-priori information input to the detector $\mathbf{I}'_{A,DET}$ according to  

    \vspace{-4mm}
    \begin{align}
      \mathbf{I}'_{A,DET}[i,k] = J \left( \sqrt{ d_{v,i} }\cdot J^{-1} ( \mathbf{I}_{A,VND}[k]) \right).
    \end{align}
    \vspace{-3mm}

  \item Compute $\mathbf{I}_{E,VND}$ according to (\ref{eqn:vnd_curve_overall})

    \vspace{-4mm}
    \begin{align}
      \mathbf{I}_{E,VND}[k] = \sum_{i=1}^D b_i \cdot J \left(\sqrt G \right)
    \end{align}
    \vspace{-3mm}

    where 

    \vspace{-4mm}
    \begin{align}
      G =&  (d_{v,i} - 1 ) J^{-1}(\mathbf{I}_{A,VND}[k]) + ...\nonumber \\ 
      &[J^{-1}(f_{DET}(\mathbf{I}'_{A,DET}[i,k]))]^2.
    \end{align}
    \vspace{-3mm}

  \end{itemize}

\end{enumerate}

The check node characteristic is completely determined by the parameters of the code and is not dependent on the properties of the detector output. Noting that $I_{E,CND} = I_{A,VND}$ compute the a-priori CND characteristic according to (\ref{eqn:ie_cnd})

\vspace{-4mm}
\begin{align}
  \mathbf{I}_{A,CND}[k] = J \left( \frac{ J^{-1}( 1 - \mathbf{I}_{E,CND}[k] ) }{ \sqrt{ d_c - 1 } } \right)
\end{align}
\vspace{-3mm}

\begin{table}[!hb]

  \centering

  \begin{tabular}{|c|c|c|c|c|}
    \hline

       Base       &     &     &     &                       \\
     Standard & $M$ & CSI & $V$ & $\frac{E_b}{N_0}$ fit \\

    \hline

    DVB-S2 & 4 &Partial  &  \scriptsize $V_1 =$ \{ 2:25920, 4:34560,  22:4320 \} & 11.9  \\
                  
    &         &          &  \scriptsize $V_2 =$\{ 2:25920, 33:4560, 30:4320 \}  & 12             \\

    &         &          & \scriptsize $V_3 =$\{ 2:25920, 4:37152, 49:1728 \} & 12             \\
    \cline{3-5}
    &         & None     &\scriptsize $V_4 =$\{ 2:25920, 3:34560, 30:4320 \} & 12.3     \\

    &         &          & \scriptsize $V_5 =$\{ 2:25920, 4:34560, 22:4320 \}  & 12.3  \\

    &         &          & \scriptsize $V_6 =$\{ 2:25920,  4:37152,  49:1728 \}  & 12.4     \\

    \cline{2-5}

    &  8 &  Partial &\scriptsize $V_7 =$\{ 2:25920,  3:34560,  30:4320 \}  & 9.5   \\

    &           &         &\scriptsize $V_8 =$\{ 2:25920,  35:3650,  3:35230 \}  & 9.5   \\

    &           &         &\scriptsize $V_9 =$\{ 2:25920,  4:34560,  22:4320 \}  & 9.7   \\

    \cline{3-5}
    &          &  None & \scriptsize $V_{10} =$\{ 2:25920,  3:34560,  30:4320 \}  & 9.8   \\

    &         &         & \scriptsize $V_{11} =$\{ 2:25920,  35:3650,  3:35230 \}  & 9.9   \\

    &         &         & \scriptsize $V_{12} =$\{ 2:25920,  4:34992,  24:3888 \}  & 10.1  \\

    \hline

    WiMax & 4  & Partial & \scriptsize $V_{13} =$\{ 2:672, 3:96, 3:1296, 9:240 \}&  12.9 \\

    &         &        & \scriptsize $V_{14} =$\{ 2:672, 3:96, 3:1356, 11:180 \} &  12.9    \\

    &         &        &   \scriptsize $V_{15} =$\{ 2:672, 3:96, 3:1376, 12:160 \} &  12.9 \\
    \cline{3-5}
     &     &  None &\scriptsize $V_{16} =$\{ 2:672,  3:96,  3:1248,  8:288 \} &  13.3  \\

    &      &       &\scriptsize $V_{17} =$\{ 2:672,  3:96,  3:1296,  9:240 \} &  13.3  \\

    &      &       &\scriptsize $V_{18} =$\{ 2:672,  3:96,  3:1356,  11:180 \} & 13.3  \\

    \cline{2-5}

    &  8 &  Partial & \scriptsize $V_{19} =$\{ 2:672,  3:96,  3:1296,  9:240 \}  & 10.4   \\

    &         &        &\scriptsize $V_{20} =$\{ 2:672,  3:96,  3:1356,  11:180 \}  & 10.4   \\

    &         &        &\scriptsize $V_{21} =$\{ 2:672,  3:96,  3:1376,  12:160 \}  &  10.4   \\
       \cline{3-5}
    &         &  None &\scriptsize $V_{22} =$\{ 2:672,  3:96, 3:1356,  11:180 \}  &  10.8   \\

    &         &        &\scriptsize $V_{23} =$\{ 2:672,  3:96,  3:1376,  12:160 \}  &  10.8   \\

    &         &        &\scriptsize $V_{24} =$\{ 2:672,  3:96,  3:1392,  13:144 \}  &  10.8  \\

    \hline

  \end{tabular}

  \caption{Degree Optimization Results}

    \vspace{-10mm}

  \label{tbl:exit_opt}

\end{table}

\subsection{Optimization Results}

In this subsection we apply the optimization procedure to improve
performance of standardized DVB-S2 and WiMax LDPC codes.
The variable node degree distributions for two standardized codes
are optimized, and their error rate performance is compared under
simulation in the DNC relay reception phase.

 \begin{figure}[!t]
   \centering
   \includegraphics[width=8cm]{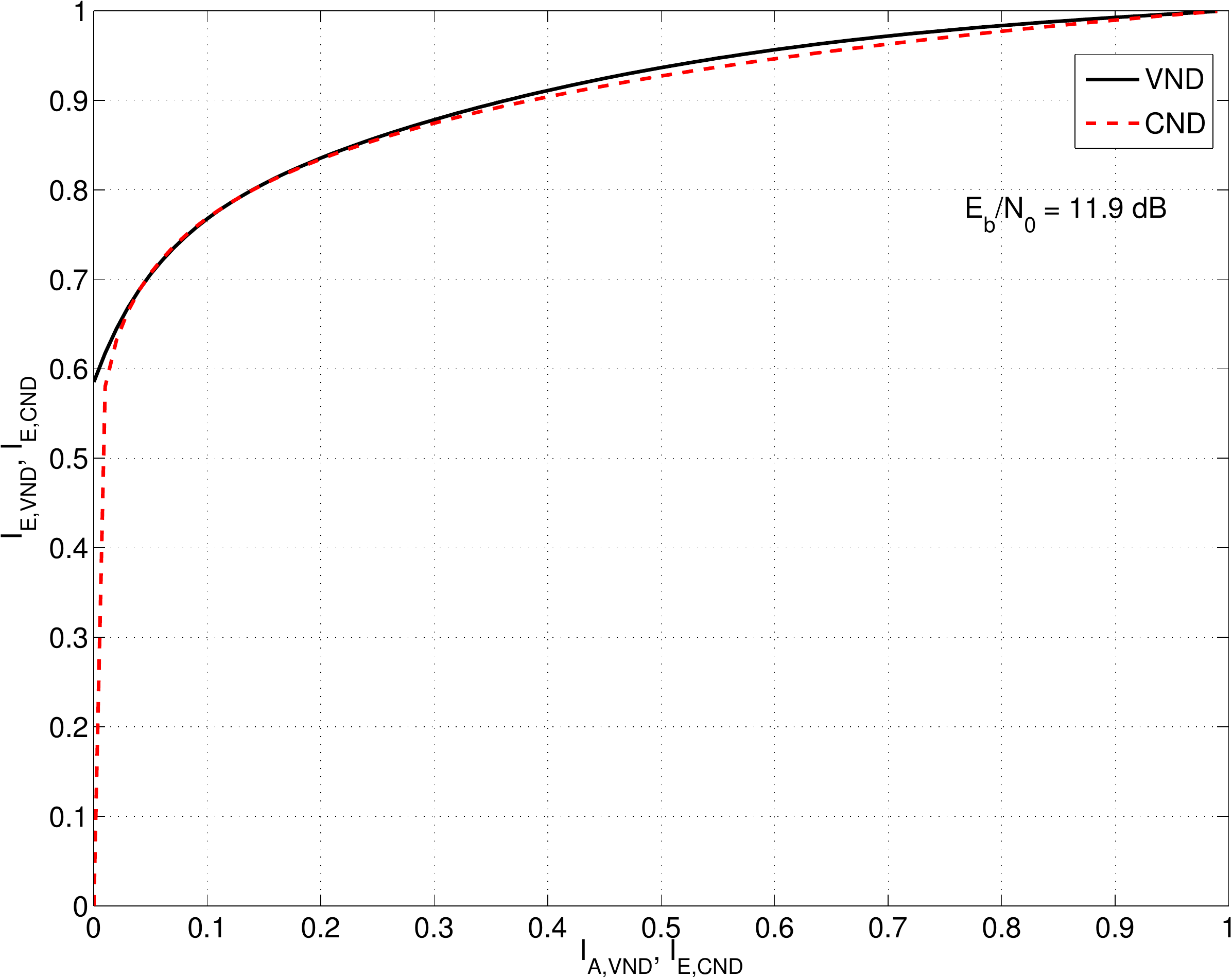}
   \caption{Example EXIT fit - DVB-S2 constraint}
   \label{fig:dvbs2exitm4c1}
 \end{figure}

\begin{figure}[b]
  \centering
  \includegraphics[width=8.8cm]{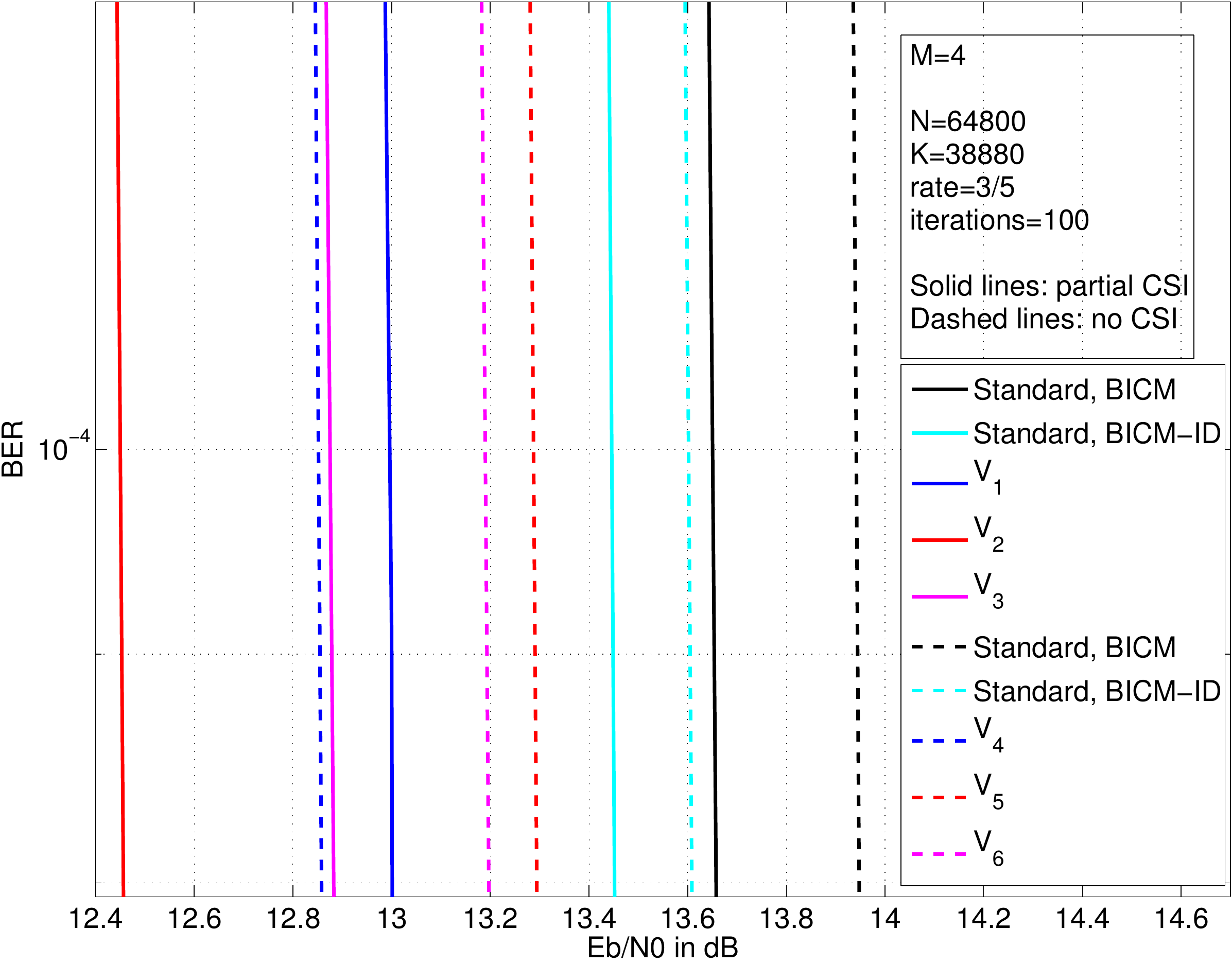}
  \caption{Bit Error Rate - DVB-S2 constrained, EXIT-optimized codes, M=4}
  \label{fig:dvbs2m4}
\end{figure}

The first code is taken from the DVB-S2 standard \cite{dvbs2:2013}, having codeword
length $N=64800$ and rate $R=3/5$.  The second code is specified
by the WiMax standard \cite{80216:2012}, having length $N=2304$ and rate $R=2/3$.
The parity check matrices for these codes are specified in their respective standards.
Note that the WiMax standard specifies two distinct parity check matrices for rate $2/3$ codes, denoted
as \emph{2/3A} and \emph{2/3B}. 
 We consider case \emph{2/3A}.
These codes satisfy the eIRA constraint and are check-regular, with $d_c = 11$ for
the DVB-S2 code and $d_c = 10$ for WiMax.

The parity check matrix for each standard codes contains a sub-matrix
satisfying the \emph{extended irregular repeat-accumulate} (eIRA) constraint.
This constraint simplifies encoding and decoding complexity.
The submatrices comprising $\mathbf{H}$ are

\vspace{-1.5mm}
\begin{align}\label{eqn:H_part}
  \mathbf{H} = [ \mathbf{H}_1 | \mathbf{H}_2 ]
\end{align}
\vspace{-4mm}

\noindent where $\mathbf{H}_2$ satisfies the eIRA constraint
and has $(N-K)$ rows and $(N-K)$ columns.
To preserve the complexity advantages of eIRA, we retain $\mathbf{H}_2$ exactly as specified
by the standards and consider optimizating the variable node degrees specifying $\mathbf{H}_1$.
Retaining $\mathbf{H}_2$ places constraints on $V$ such that codes based on DVB-S2
have $d_{v,1} = 2, o_1 = 25920$, and codes based on WiMax have $d_{v,1} = 2, o_1 = 672$
and $d_{v,2} = 3, o_2 = 96$.
All other degrees may be chosen freely.

\begin{figure}[!b]
  \centering
  \includegraphics[width=8.8cm]{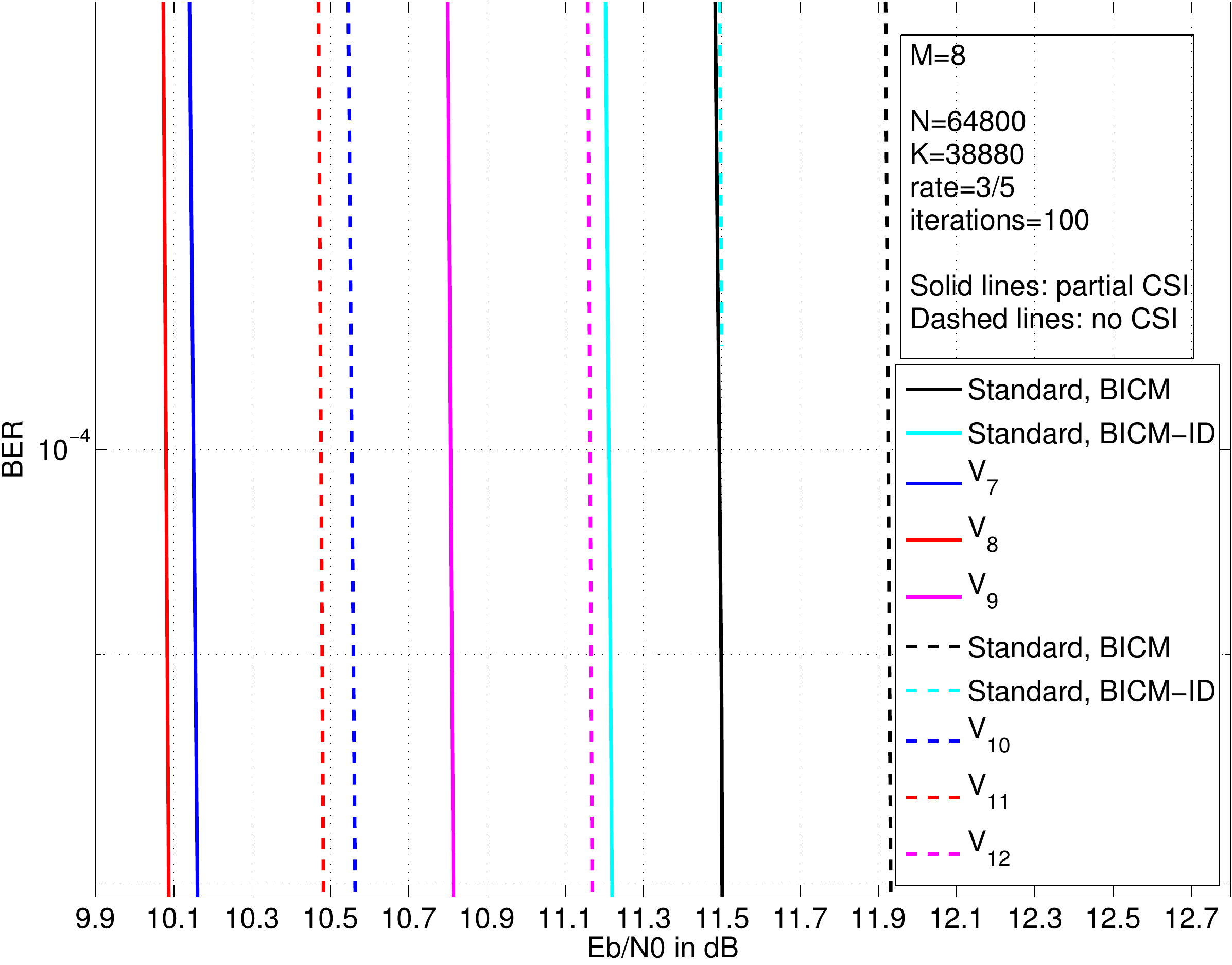}
  \caption{Bit Error Rate - DVB-S2 constrained, EXIT-optimized codes, M=8}
  \label{fig:dvbs2m8}
\end{figure}

Variable node degree optimization is performed under several cases
of receiver configuration and channel state information, specifically,
modulation orders $M={ 4 , 8 }$ with and without CSI, for a total
of four configurations.
A range of degree distributions is considered for each code and receiver configuration, and 
the best performing degree distribution of each is realized and simulated.

Considering the DVB-S2 LDPC code, the number of distinct variable node degrees is
$D=3$, and the degrees considered are all combinations of
the sets $d_{v,1} = 2$, $d_{v,2} \in \{ 3, 4, ... , 100 \}$, 
and $d_{v,3} \in \{ d_{v,2}+1, d_{v,2}+2, ..., 100 \}$ which satisfy the constraints
for realizable codes described in Subsection \ref{subsec:opt_vnd}.
Considering the WiMax code, the number of distinct variable node degrees is $D=4$,
and the degrees considered are all possible combinations of $d_{v,1} = 2$,
 $d_{v,2} = 3$, $d_{v,3} \in \{ 1, 2, ... , 100 \}$, and $d_{v,4} \in \{1, 2, ..., 100 \}$.

For every receiver configuration and degree distribution $V$, the EXIT threshold is computed to discover
the the single variable node degree distribution having the best performance.
For each distribution, the threshold is defined as the highest SNR value $\frac{E_b}{N_0}$ for which the VND and CND decoder EXIT curves touch. 
The best performing distribution is defined as minimizing this SNR value.
An example is shown in Fig. \ref{fig:dvbs2exitm4c1}.

The optimization results are shown in Table \ref{tbl:exit_opt}.
For each degree distribution $V$, a code is realized having parity check matrix
satisfying the degree distribution and maintaining the eIRA constraint.
The parity check matrix is generated randomly.
In the interest of space, further details of parity check matrix generation are omitted here.

\subsection{Error-rate Performance of Optimized Codes}

The error-rate performance of the optimal randomly generated codes is compared
to the standardized codes via Monte Carlo simulation according to the model
described in Section \ref{sec:SystemModel}.

The source nodes generate information sequences having length $K=38800$ for the DVB-S2 codes and $K=1536$ for WiMax.
The information sequences are channel encoded to produce codewords having length $N=64800$ for DVB-S2
and $N=2304$ for WiMax.
These codewords are mapped to FSK symbols.
Modulation orders $M=\{4,8\}$ are considered.
The modulated symbols are transmitted over the channel and received at the relay.
The relay performs decoding considering the cases of having channel state information
and no channel state information.

\begin{figure}[]
  \centering
  \includegraphics[width=8.8cm]{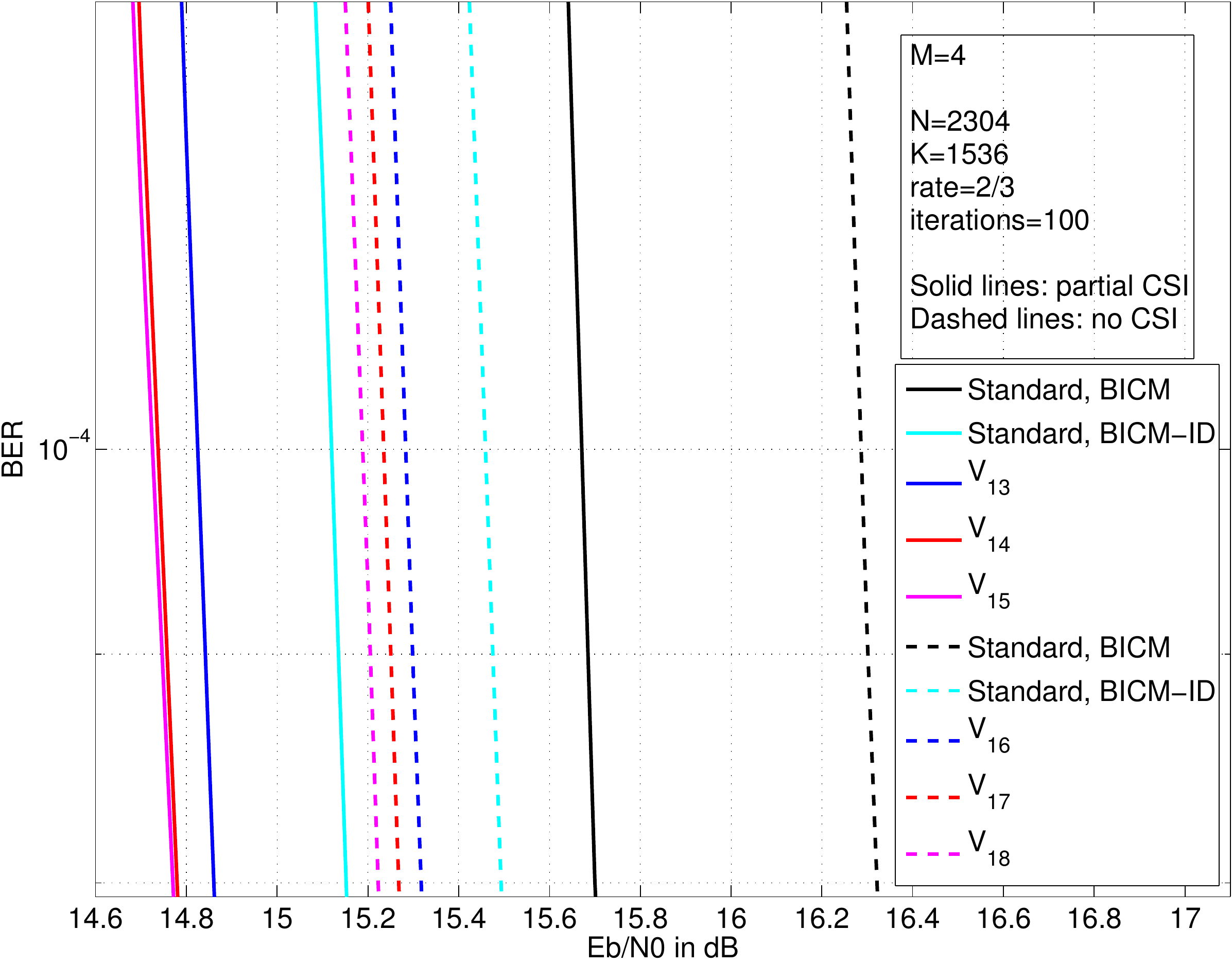}
  \caption{Bit Error Rate - WiMax constrained, EXIT-optimized codes, M=4}
  \label{fig:wimaxm4}
\end{figure}

\begin{figure}[]
  \centering
  \includegraphics[width=8.8cm]{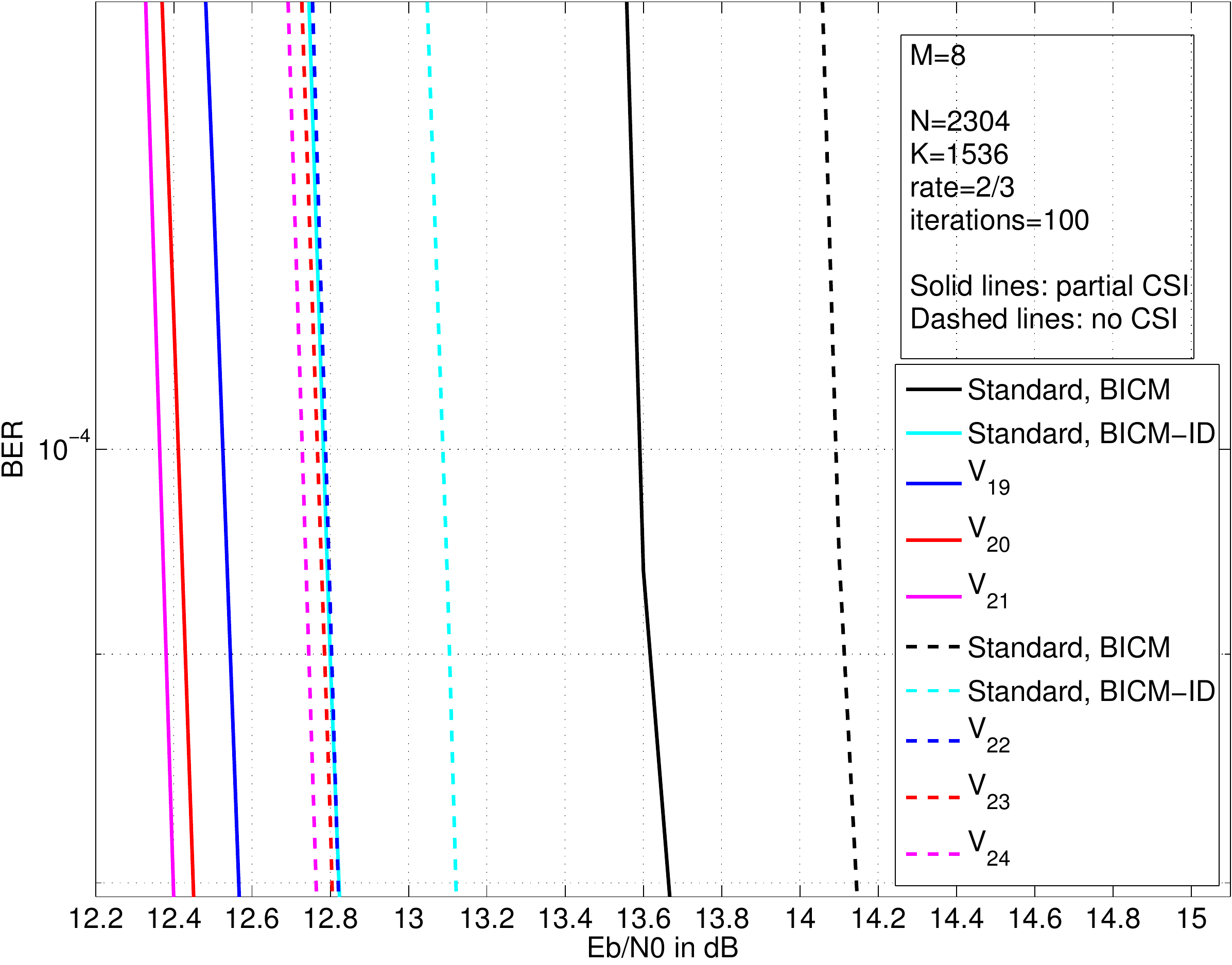}
  \caption{Bit Error Rate - WiMax constrained, EXIT-optimized codes, M=8}
  \label{fig:wimaxm8}
\end{figure}

In all cases the relay receiver performs $100$ decoding iterations before
making a hard decision on the network-coded bits.
The standard codes are simulated for two cases of decoder-to-demodulator
extrinsic information feedback:
BICM, in which no information is fed back from decoder to demodulator,
and BICM-ID, in which information is fed back to the demodulator after each decoding iteration.
Simulation of the randomly generated codes always uses BICM-ID.

Error rate performance for DVB-S2 and WiMax codes, $M=\{4,8\}$,
and for partial and no CSI are shown in Figures \ref{fig:dvbs2m4}, 
\ref{fig:dvbs2m8}, \ref{fig:wimaxm4} and \ref{fig:wimaxm8}.
 The degree distributions for each plotted code are denoted on the legends
as $V_i$, referring to a degree distribution listed in the column $V$ in Table \ref{tbl:exit_opt}.

Consider the results for DVB-S2.
For $M=4$ and partial CSI, the best performing optimized code shows
an energy efficiency improvement of approximately $1$ dB, while
the no CSI case shows improvement of roughly $0.8$ dB.
For $M=8$ and partial CSI, the improvement is about $1.2$ dB,
while no CSI shows $1$ dB of improvement.

Now consider WiMax results.
For $M=4$ and partial CSI, performance gain of the best optimized code
over the standard code is about $0.4$ dB, while for the no CSI case, improvement is about $0.3$ dB.
For $M=8$ and partial CSI, improvement is $0.4$ dB,
while no CSI shows $0.3$ dB of improvement.

\section{Conclusion}

 This work has presented an optimization procedure for LDPC
 Tanner Graph variable node degree distribution in the 
 physical-layer network coded two way relay channel.
 The codes are optimized for relay reception in the multiple access phase.
 LDPC Codes from the DVB-S2 and WiMax standards are used to establish
 baseline error-rate performance.
 Optimized codes are generated under a variety of system configurations:
 modulation order $M=4$ and $M=8$, 
 The optimized codes outperform the standard codes by margins from $0.3$ to
 $1.2$ dB.

\balance

\bibliographystyle{IEEEtran}
\bibliography{bibliography}

\end{document}